%% file: main.tex
\documentstyle[12pt,psfig]{article}
\def\ro{{\rm o}}
\def\fos{freeze-out shock\,}
\def\D{{\Delta}}
\def\lp{\left(}
\def\rp{\right)}
\begin{document}
\begin{titlepage}
\title{ Particle freeze-out and discontinuities\\
in relativistic hydrodynamics}

\vspace{0.5cm}
 
\author{
K A Bugaev$^{\dagger \ddagger }$, M I Gorenstein$^{\dagger \ddagger}$
and  W Greiner$^{\dagger }$ \\
\\
\\
$^{\dagger}$ Institute for Theoretical Physics, Goethe University,
Frankfurt, Germany\\
$^{\ddagger}$ Bogolyubov Institute for Theoretical Physics,
Kiev, Ukraine
}

\date{}
\maketitle

\vspace{1cm}
{\bf Abstract.}~~~
Freeze-out of particles in relativistic hydrodynamics 
is considered across a 
3-dimensional space-time hypersurface. 
The conservation laws for time-like parts of the freeze-out
hypersurface 
require different values of temperature, baryonic 
chemical potential and flow velocity 
in the fluid and in the final particle spectra.
We analyze this freeze-out discontinuity and its connection to 
the shock-wave phenomena in relativistic hydrodynamics.

\end{titlepage}

%\newpage
\noindent
{\bf 1. Introduction}

\vspace{0.3cm}
\noindent
Relativistic hydrodynamical model \cite{Lan53} has
been widely discussed in recent years within their connection to
high energy nucleus-nucleus (A+A) collisions (see, for example, 
\cite{rev}). This approach is the most straight and unique way
to investigate the hadron-matter equation of state
(EoS) and possible phase transition into a deconfined state
called the quark-gluon plasma. 
The aim of the relativistic
hydrodynamical model of A+A collisions is to elucidate 
the properties of the highly excited matter comparing the model results
with experimental observables.  
The system evolution in relativistic hydrodynamics is governed by
the energy-momentum tensor
\begin{equation}\label{tmunu}
 T^{\mu\nu}~ =~ (\varepsilon + p)u^{\mu}u^{\nu}~ -~ pg^{\mu\nu}
\end{equation}
and conserved charge currents. 
The baryonic current,
\begin{equation}\label{barcur}
 j^{\mu}~ =~ n~u^{\mu}~,
\end{equation}
plays the main role in the applications to A+A collisions.
The hydrodynamical description includes 
the local thermodynamical variables
(energy density $\varepsilon$, pressure $p$,
baryonic density $n$) and the collective
four-velocity $u^{\mu}=(1-{\bf v}^2)^{-1/2}(1, {\bf v})$.
The continuous flows are the solutions of the
hydrodynamical equations
\begin{equation}\label{hydroeq}
\partial _{\mu}T^{\mu\nu}~ =~0~,~~~~~
\partial _{\mu}j^{\mu}~ =~0~,
\end{equation}
with specified initial conditions.
To complete the system of hydrodynamical equations (\ref{hydroeq})
the fluid EoS has to be added as an input
$p=p(\varepsilon ,n)$.
It is usually used in a more convenient  
form of $p=p(T,\mu_B)$ with temperature
$T$ and baryonic chemical potential $\mu_B$ as independent
variables. Other thermodynamical functions can be found 
from the following thermodynamical identities
($s$ is the entropy density):
\begin{equation}\label{thermiden}
s~=~\left(\frac{\partial p}{\partial T}\right)_{\mu_B}~,~~~~
n~=~\left(\frac{\partial p}{\partial \mu_B}\right)_{T}~,~~~~
\varepsilon~=~Ts+\mu_B n -p~.
\end{equation}

Besides the fluid EoS and the initial conditions 
for the fluid evolution
the third crucial ingredient of the hydrodynamical model 
of A+A collisions
is the so-called freeze-out (FO) procedure, i.e. the prescription
for the calculations of the final hadron observables.  
Particles which leave
the system and reach the detectors are considered 
via FO scheme, 
where the frozen-out particles are formed 
on a 3-dimensional hypersurface in space-time.  
The FO
description has a straight influence on particle momentum spectra
and therefore on all hadron observables.
During last three years an essential progress
has been achieved in the formulation 
of the FO procedure in relativistic hydrodynamics.
A generalization of the well known Cooper-Frye (CF)
formula \cite{CF74} to the case of time-like 
FO hypersurface was suggested in Ref.~\cite{Bu96}.
The new formula does not contain negative particle number
contributions on time-like FO hypersurface appeared 
in the CF procedure from those particles which cannot leave
the fluid during its hydrodynamical expansion.
The FO procedure of Ref.~\cite{Bu96} has been further developed
in a series of publications [5-10]. 
Ref. [8] presents also the kinetic approach to the FO problem. 

There is a new distinct and unusual feature of the proposed FO procedure.
The parameters of the distribution function
for the final particles should be different from those in the fluid
to satisfy the conservation laws.
The particle emission from the time-like parts of the FO hypersurface looks 
as a `discontinuity' in the hydrodynamic motion. We call it the FO shock.
In  the present paper we analyze a connection of
this new type of discontinuities to 
a general problem of shock waves 
in relativistic hydrodynamics. 

\vspace{0.5cm}
\noindent
{\bf 2. Freeze-out procedure}

\vspace{0.3cm}
\noindent
In the hydrodynamical model of A+A collisions the fluid expansion 
has to be ended by the correct transformation
of the fluid into free streaming
hadrons. The most important requirement for this FO procedure
is to satisfy energy-momentum and charge conservations
during the fluid transition into free final particles.
It is assumed that there is a narrow space-time region where
the mean free path of the fluid constituents increases
rapidly and becomes comparable with the characteristic
size of the system. The local thermal equilibrium is supposedly
maintained in this intermediate region. In practice, one considers
a ``zero width'' approximation and introduces a FO surface,
so that particle distributions remain frozen-out from there on.

In his original paper Landau \cite{Lan53} defined the
FO hypersurface by the condition \mbox{$T(t,x)=T^*$}, where
$t$ and $x$ are the time and the space coordinate respectively, and $T^*$ 
is the fluid FO temperature chosen to be approximately equal to the
pion mass. 
The FO procedure, first introduced by Milekhin \cite{Mi58}, was 
based on the intuitive physical 
assumption of isotropic thermal particle spectra in the rest
frames of each fluid element. The final hadron
observables was then obtained by  the Lorentz transformations
of these spectra 
with known hydrodynamical velocities and integration over all
fluid elements.   
The FO procedure was improved further by Cooper and Frye \cite{CF74} and
this method has been used ever since. 
Final spectrum for {\it i}-th type hadron is expressed by formula 
\cite{CF74}:
\begin{equation}\label{CF}
k^{\ro} \frac{d^3N_i}{dk^3}~=
~\frac{d_i}{(2\pi)^3}\int_{\Sigma_f}d\sigma_{\mu}k^{\mu}~
\phi_i\left(\frac{k^{\nu}u_{\nu} - \mu_i}{T^*}\right)~,
\end{equation}
where $d\sigma_{\mu}$ 
is the external normal 4-vector to the FO 
hypersurface $\Sigma_f$,   $\phi_i$ denotes the local thermal
Bose or Fermi distributions, $d_i$ is the degeneracy factor for
particle $i$ with the chemical potential $\mu_i$.
The particle spectrum (\ref{CF}) is known as the Cooper-Frye (CF)
distribution function. 
The chemical potential $\mu_i$ is equal to
\begin{equation}\label{chempot}
\mu_i~=~b_i \mu_B~+~ s_i\mu_S ~+~q_i\mu_Q~,
\end{equation}
where $b_i,s_i,q_i$ are the baryonic number, strangeness
and electric charge of $i$-th particle. The strange chemical potential 
$\mu_S$ and electric chemical potential $\mu_Q$ are always   
chosen to make the total strangeness to be equal to zero
and the ratio of electric charge to the baryonic number
to be fixed at their initial value given by A+A collision.
Therefore, only the baryonic chemical potential $\mu_B$
is considered as independent 
thermodynamical variable.

The CF formula (\ref{CF})
corrected the Milekhin's approach. The matter is
that the external normal
vector $d\sigma_{\mu}$ equal to $(-dx,dt)$ in 1+1 dimensions
does not in general become
equal to $d\sigma^*_{\mu}=(-dx^*,0)$
after the Lorentz transformation into the fluid element rest frame.
The non-simultaneous particle FO in the rest frame   
of the fluid element modifies the thermal distribution
function and causes a nonzero momentum of the fluid element
in its rest frame (``local anisotropy'', see Ref.~\cite{Gor84}
for details).

The initial conditions of the hydrodynamical motion
are given on the initial hypersurface $\Sigma_{in}$.
The final hypersurface
$\Sigma_{f}$ should be closed to $\Sigma_{in}$ and 
in general $\Sigma_{f}$ consists of both space-like (s.l.) 
and time-like (t.l.) parts. Using the Gauss theorem
and Eq.~(\ref{hydroeq}) one can easily 
prove the energy-momentum and charge number conservations
with CF FO formula:
\begin{equation}\label{CFconserv}
\int_{\Sigma_{in}}d\sigma_{\mu}~T^{\mu\nu}~=~
\int_{\Sigma_{f}}d\sigma_{\mu}~T^{\mu\nu} ~\equiv~
\sum_i \frac{d_i}{(2\pi)^3}\int_{\Sigma_{f}}d\sigma_{\mu}~
\int\frac{d^3k}{k^\ro}~k^{\mu}k^{\nu}~
\phi_i\left(\frac{k_{\lambda}u^{\lambda} -\mu_i}{T^*}\right)~,
\end{equation}
\begin{equation}\label{CFconserv1}
\int_{\Sigma_{in}}d\sigma_{\mu}~j^{\mu}~=~
\int_{\Sigma_{f}}d\sigma_{\mu}~j^{\mu}~\equiv~
\sum_i \frac{d_i b_i}{(2\pi)^3}\int_{\Sigma_{f}}d\sigma_{\mu}~
\int d^3k ~k^{\mu}~
\phi_i\left(\frac{k_{\lambda}u^{\lambda} -\mu_i}{T^*}\right)~,
\end{equation}
where $b_i$ is the baryonic number of $i$-th particle
and the summation over
all particles and the corresponding antiparticles
is assumed in
the right hand side of Eqs.~(\ref{CFconserv},\ref{CFconserv1}).
Note that fluid EoS at the FO is assumed to be an
ideal gas EoS for a mixture
of different particle species. The hydrodynamical quantities
$T^{\mu\nu}$ (\ref{tmunu}) and $j^{\mu}$ (\ref{barcur})
are identically presented on the FO surface $\Sigma_f$
in terms of final particle distributions $\phi _i$
according to Eqs.~(\ref{CFconserv},\ref{CFconserv1}). 

The CF formula (\ref{CF}) still does
not provide a complete solution of the FO problem.
The FO surface consists of t.l. parts too.
Eq.~(\ref{CF}) can not however be used for a t.l. FO surface
(i.e., s.l. normal vector $d\sigma_{\mu}$):
free final particles ``return'' to the fluid if $d\sigma_{\mu}k^{\mu} < 0$,
and this causes unphysical {\it negative} contributions to the
{\it number} of final particles. 
A first attempt to improve the CF formula for t.l.
FO hypersurface was done by Sinyukov in Ref.~\cite{Si:89}.
The modified
FO procedure and new formula for the final particle spectra 
emitted from the t.l. FO hypersurface was proposed in Ref.~\cite{Bu96}:
\begin{equation}\label{CO}
k^{\ro} \frac{d^3N_i}{dk^3}~= 
~\frac{d_i}{(2\pi)^3}\int_{\Sigma_f}d\sigma_{\mu}k^{\mu}~
\phi_i\left(\frac{k_{\lambda}u^{\lambda}_g - 
\mu_i^g}{T_g}\right)~\theta (d\sigma_{\nu}k^{\nu})~. 
\end{equation}
Eq.~(\ref{CO}) looks like CF formula (\ref{CF}), but without negative
particle numbers that appear in (\ref{CF}) for t.l. FO surfaces.
These negative contributions are cut-off by the $\theta$-function
in Eq.~(\ref{CO}). We'll call Eq.~(\ref{CO}) the cut-off (CO)
distribution in what follows. 
An inclusion of the CO FO into the self-consistent 
hydrodynamical scheme is considered in Ref.~\cite{6}.
The distribution function
for the free particles in Eq.~(\ref{CO}) contains the new parameters
$T_g, \mu_i^g, u^{\nu}_g$. 
We'll briefly call the free particle state as the `gas'
%%%
in order to distinguish it from the ordinary fluid.
Note, however, that final particle system differs from the normal 
fluid and gas.
It has the non-thermal CO distribution function 
which is frozen-out as 
all further particle rescatterings are `forbidden' in the
post-FO gas state. 

The conservation laws 
across FO hypersurface $\Sigma_f$ can be written now in their local form as
\begin{eqnarray}\label{BGconserv} 
& &
\sum_i \frac{d_i}{(2\pi)^3}
%\int_{\Sigma_{f}}
d\sigma_{\mu}~
\int\frac{d^3k}{k^\ro}~k^{\mu}k^{\nu}~
\phi_i\left(\frac{k_{\lambda}u^{\lambda}_f -\mu_i^f}{T_f}\right)
%~ = ~
\\ &=&~
\sum_i \frac{d_i}{(2\pi)^3}
%\int_{\Sigma_{f}}
d\sigma_{\mu}~
\int\frac{d^3k}{k^\ro}~k^{\mu}k^{\nu}~
\phi_i\left(\frac{k_{\lambda}u^{\lambda}_g -\mu_i^g}{T_g}\right)~
\theta (d\sigma_{\nu}k^{\nu})~, 
\nonumber
\end{eqnarray}
\begin{eqnarray}\label{BGconserv1}
%\begin{equation}\label{BGconserv1}
& &
\sum_i \frac{d_i b_i}{(2\pi)^3}
%\int_{\Sigma_{f}}
d\sigma_{\mu}~ 
\int d^3k ~k^{\mu}~
\phi_i\left(\frac{k_{\lambda}u^{\lambda}_f -\mu_i^f}{T_f}\right)
%~=~
\\
&= &~
\sum_i \frac{d_i b_i}{(2\pi)^3}
%\int_{\Sigma_{f}}
d\sigma_{\mu}~
\int d^3k ~k^{\mu}~
\phi_i\left(\frac{k_{\lambda}u^{\lambda}_g -\mu_i^g}{T_g}\right)~
\theta (d\sigma_{\nu}k^{\nu})~. 
\nonumber
\end{eqnarray}
These conservation laws  were derived together with the equation of motion of 
the fluid in Ref. \cite{6}
and should be solved simultaneously with the latter ones 
in order to find the FO hypersurface in the practical hydrodynamical
calculations.
%Here we, however, analyze its most important aspects only. 

The presence of the $\theta$-function in the right hand side of 
Eqs.~(\ref{BGconserv},\ref{BGconserv1}) leads to the discontinuity
between the fluid 
$T_f, \mu_i^f, u^{\nu}_f$ 
and the gas 
$T_g, \mu_i^g, u^{\nu}_g$
variables 
to satisfy the energy-momentum and baryonic number conservation.
We call this discontinuity the FO shock. 
Note that the strangeness flow across the FO surface is equal to zero
and electric charge flow is proportional to the baryonic number flow.
Therefore, the conservation of strangeness and electric charge is
also fulfilled.

Another FO discontinuity  problem occurs in the hydrodynamical
model calculations when the EoS is 
different on the two 
sides of the FO front (for the final gas of free particles one must use
the ideal gas distributions). The conservation laws 
on the FO surface in this case should lead to the discontinuities between 
fluid and gas of final particles across both t.l.
and s.l. parts of $\Sigma _f$  
\cite{Go90}.
The discontinuity across the s.l. part of $\Sigma _f$ corresponds to the 
so-called time-like shocks \cite{Cse87}
(i.e., $d\sigma_{\mu}$ is t.l. vector) and will be also discussed in 
the next Sections.

\vspace{0.5cm}
\noindent
{\bf 3. Relativistic shock waves}

\vspace{0.3cm}
\noindent
Equations (\ref{hydroeq}) 
are the differential form
of the energy-momentum and baryonic number conservation laws. Along
with these continuous flows, the conservation laws can also be
realized in the form of discontinuous hydrodynamical flows which are
called shock waves and satisfy the following equations:
\begin{equation}\label{enmombar}
 T_\ro^{\mu\nu}\Lambda_{\nu} = T^{\mu\nu}\Lambda_{\nu}  \;\; ,~~~~
% \end{equation}
% \begin{equation}
 n_{\ro}u_{\ro}^{\mu}\Lambda_{\mu}= n u^{\mu}\Lambda_{\mu} \;\; ,
\end{equation}
where $\Lambda _{\mu}$ is the unit 4-vector normal to the
discontinuity hypersurface. 
In Eq.~(\ref{enmombar}) 
the zero subscript corresponds
to the initial state ahead of the shock front and the quantities without
an index are the final state values behind it.

The important constraint on the transitions 
(\ref{enmombar}) 
is the requirement of non-decreasing
entropy
(thermodynamical
stability condition):
 \begin{equation}\label{entropy}
su^{\mu}\Lambda _{\mu} \ge s_{\ro}u_{\ro}^{\mu}\Lambda _{\mu} \; .
\end{equation}

We consider one-dimensional hydrodynamical motion in what
follows. In its usual sense
the theory of the shock waves corresponds to
the discontinuities on the surface with 
space-like $\Lambda _{\mu}$, i.e., the shock-front velocity
is smaller than 1.  
In this case one can always choose
the Lorentz frame where the shock front is at rest. Then the
hypersurface of shock discontinuity is
$x_{sh} = const$ and $\Lambda _{\mu} =(0,1)$. The shock
equations (\ref{enmombar}) 
%in this traditional case (we call them s.l. shocks) are:
become 
\begin{equation}\label{slshock}
T_{\ro}^{01} = T^{01}, \; \; T_{\ro}^{11} = T^{11} \;\; ,~~~
n_{\ro}u_{\ro}^{1} = nu^{1} \; \; .
\end{equation}
From Eq. (\ref{slshock}) one obtains
\begin{equation}\label{slvel}
v_{\ro}^{2} = \frac{(p - p_{\ro})(\varepsilon + p_{\ro})}
{(\varepsilon - \varepsilon_{\ro})(\varepsilon_{\ro} + p)} \; , \; \;
v^{2} = \frac{(p - p_{\ro})(\varepsilon_{\ro} + p)}
{(\varepsilon - \varepsilon_{\ro})(\varepsilon + p_{\ro})} \;\; .
\end{equation}
Substituting (\ref{slvel}) into the last equation in (\ref{slshock}) 
we obtain the well known
Taub adiabate (TA) equation ~\cite{Tau48}
\begin{equation}\label{TA}
n^{2}X^{2} - n_{\ro}^{2}X_{\ro}^{2} - (p - p_{\ro})
(X + X_{\ro}) = 0 \;\; ,
\end{equation}
in which $X \equiv (\varepsilon + p)/n^{2}$, and TA therefore
contains only the thermodynamical variables.

Eq.~(\ref{TA}) can be written also in the form
\begin{equation}\label{nTA}
n^2~=~n_{\ro}^{2}~ \frac{(\varepsilon + p_{\ro})(\varepsilon + p)}
{(\varepsilon_{\ro} + p_{\ro})(\varepsilon_{\ro} + p)}~.
\end{equation}
If the equation of state $p=p(\varepsilon, n)$ is given,
then Eq.~(\ref{nTA}) defines the function $n=n_T(\varepsilon)$
which is the TA (\ref{TA}) in the $(\varepsilon, n)$ plane.
The TA in the $(\varepsilon, p)$ plane is defined by the function
$p_T(\varepsilon) = p(\varepsilon, n_T(\varepsilon))$.
In the case of zero baryonic number it is just reduced
to the EoS $p_T(\varepsilon)=p(\varepsilon)$.
The point $(\varepsilon_{\ro}, p_{\ro},n_{\ro})$ is called the center of 
the TA.
The shock transition from the state $(\varepsilon_{\ro}, p_{\ro}, 
n_{\ro})$ to $(\varepsilon_1, p_1, n_1)$ is mechanically stable 
if for all $\varepsilon$ between $\varepsilon_{\ro}$
and  $\varepsilon_1$ the following inequality is valid
\cite{Go87}:
\begin{equation}\label{mechstab} 
(\varepsilon_1 - \varepsilon_{\ro})~ \left[p_{cr}(\varepsilon ) -
p_T(\varepsilon)\right]~\geq~ 0,
\end{equation}
where
\begin{equation}\label{pcr}
p_{cr}(\varepsilon)~=~A~-~\frac{C}{A+ \varepsilon}
\end{equation}
with
\begin{equation}\label{AC}     
A~=~\frac{\varepsilon_1p_1 - \varepsilon_{\ro}p_{\ro}}
{\varepsilon_1 -\varepsilon_{\ro}+p_{\ro}-p_1 }~,~~~
C=(A+\varepsilon_{\ro})(A-p_{\ro})=
(A+\varepsilon_1)(A-p_1)~.
\end{equation}
Thermodynamical stability (\ref{entropy}) follows from the mechanical
stability (\ref{mechstab}), and the inverse statement
is not in general true \cite{Go87}.
It follows from (\ref{mechstab}) that the necessary and sufficient condition
for the mechanical stability of shocks is that the compression 
($\varepsilon_1 > \varepsilon_{\ro}$) 
TA $p_T(\varepsilon)$ goes under the critical curve $p_{cr}(\varepsilon)$
(\ref{pcr}) and the rarefaction 
$(\varepsilon_1 < \varepsilon_{\ro})$
TA goes above $p_{cr}(\varepsilon)$ for all $\varepsilon$ between the 
initial $\varepsilon_{\ro}$ and final $\varepsilon_{1}$ points.
The consequences of these relations are the well-known inequalities for 
the speed of sound, 
$c_s=[(\partial p/\partial \varepsilon)_{s/n}]^{1/2}$,
and the flow velocities at both sides of the shock
front in its rest frame
\begin{equation}\label{sound}
c_{s0}~\leq ~v_0~,~~~~c_{s1}~\geq ~v_1~,
\end{equation}
which are the necessary conditions of the general mechanical 
stability criterion (\ref{mechstab}).

The function 
\begin{equation}\label{z}
z_T(\varepsilon)~=~\frac{d^2p_T}{d\varepsilon^2}
+\frac{2}{\varepsilon + p_T}~\frac{dp_T}{d\varepsilon}
\left(1-\frac{dp_T}{d\varepsilon}\right)~.
\end{equation}
defines the thermodynamic properties, i.e.,
the medium is said to be thermodynamically normal
if the quantity $z_T$
is positive, and thermodynamically anomalous when  $z_T<0$.
In the normal case $z_T>0$ the compression shocks are stable
whereas the rarefaction shocks become stable in the thermodynamically
anomalous media with $z_T<0$. Note that the function $p_{cr}(\varepsilon)$
(\ref{pcr}) is the solution of the equation $z_T(\varepsilon)=0$. 
The case of varying sign of  $z_T(\varepsilon)$ which is typical for the 
phase transitions are discussed in Refs.~\cite{Bug89,Bug91}.

Let us consider now the discontinuities on a hypersurface with a t.l. normal 
vector $\Lambda _{\mu}$. This possibility
was suggested by Csernai in Ref.~\cite{Cse87}. We call them 
t.l. shocks. In this case
one can always choose another convenient Lorentz
frame ('simultaneous system') where the hypersurface of the
discontinuity is
$t_{sh} = const$ and $\Lambda _{\nu}
= (1,0)$. Equations (\ref{enmombar}) become then
\begin{equation}\label{tlshock}
T_{\ro}^{00} = T^{00}, \; \; T_{\ro}^{10} = T^{10} \;\; ,~~~
% \end{equation}
% \begin{equation}
n_{\ro}u_{\ro}^{0} = nu^{0} \;\; .
\end{equation}
From Eq. (\ref{tlshock}) we find
\begin{equation}\label{tlvel}
\tilde{v}_{\ro}^{2} =
\frac{(\varepsilon - \varepsilon_{\ro})(\varepsilon_{\ro} + p)}
{(p - p_{\ro})(\varepsilon + p_{\ro})}~, \; \; ~~
\tilde{v}^{2} = \frac
{(\varepsilon - \varepsilon_{\ro})(\varepsilon + p_{\ro})}
{(p - p_{\ro})(\varepsilon_{\ro} + p)} \;\;,
\end{equation}
where we use $``\sim"$ sign to distinguish the t.l. shock case
(\ref{tlvel}) from the standard s.l. shocks (\ref{slvel}). 
Substituting (\ref{tlvel}) 
into the last equation in (\ref{tlshock})
one finds the equation for t.l. shocks  which is identical to
the TA  of Eq.~(\ref{TA}). We stress, however, that the intermediate steps 
%(Eqs. (\ref{tlvel})and (\ref{slvel})) 
are 
quite different. The two solutions,
Eqs.~(\ref{tlvel})
and (\ref{slvel}),
are connected to each other by simple relations,
\begin{equation}\label{tlslvel}
\tilde{v}_{\ro}^{2} =
\frac{1}{v_{\ro}^{2}}~, \; \;~~
\tilde{v}^{2} = \frac
{1}{v^{2}} \;\; ,
\end{equation}
between the velocities of s.l. and t.l. shocks. These relations
show that only one type of transitions can be realized for the given
initial  and final states.
Physical regions $[0,1)$ for $v_{\ro}^{2},v^{2}$ (\ref{slvel}) and for
$\tilde{v}_{\ro}^{2},\tilde{v}^{2}$ (\ref{tlvel}) can be easily found in
the $(\varepsilon $--$p)$-plane shown
in \mbox{Fig. 1.} 

If one takes as the initial and final states only states which
are thermodynamically equilibrated, 
the TA passes then through the center of TA $(\varepsilon _{\ro},p_{\ro})$. 
For any physical EoS with $0\leq c_s\leq 1$
TA lies as a whole in the regions 
I and IV in Fig. 1 \cite{Bug88}, i.e. only compression 
and/or rarefaction s.l. shocks 
(with s.l. normal vector $\Lambda_{\mu}$) are permitable. 
%Phase 
%transitions via the compression s.l. shocks into  region IV
%and rarefaction s.l. shocks into region I. 
%%% NEWREF

The only way to make the t.l. shocks to be possible
is to allow the metastable initial and/or final states.
The TA then no 
longer passes through the initial point $(\varepsilon _{\ro},p_{\ro})$ and
new possibilities of t.l. shock
transitions (\ref{tlshock},\ref{tlvel}) to regions III and VI in Fig. 1 
would be realized (see, e.g.,  the t.l. shock hadronization of supecooled 
quark-gluon plasma  in Ref.~\cite{Gor94}).

\vspace{0.5cm}
\noindent
{\bf 4. Freeze out shocks}

\vspace{0.3cm}
\noindent
The conservation laws (\ref{BGconserv},\ref{BGconserv1})
between the fluid and the free streaming particles are simplified across the s.l. FO 
hypersurface. In this case the $\theta$-function in the right hand side
of Eqs.~(\ref{BGconserv},\ref{BGconserv1}) is equal to unity
for all momenta of final particles. The CO formula (\ref{CO})
is reduced to the CF distribution (\ref{CF}) at  s.l. parts of the FO
hypersurface. It was shown in the previous Section that t.l. shocks
(shocks across s.l. hypersurface) are not permitable for any physical
EoS, in particular for the ideal gas EoS for the fluid and the
gas of free particles. It means that the only solution of
Eqs.~(\ref{BGconserv},\ref{BGconserv1}) is 
\begin{equation}\label{smooth}
T_f=T_g~,~~~~ \mu_i^f=\mu_i^g ,~~~~ u_f^{\mu}=u_g^{\mu}~. 
\end{equation}
No discontinuities between the fluid and the gas parameters can exist,
and the particle emission from s.l. hypersurface is a smooth
FO according to  the CF formula (\ref{CF}). 
We do not discuss here 
the possibilities when the fluid FO takes place
from the state which is out of local thermodynamical equilibrium. 

For the t.l. parts of the FO hypersurface $\Sigma_f$
a rather different solution should be
realized. The $\theta$-function in Eqs.~(\ref{BGconserv},\ref{BGconserv1})
equals to zero when $d\sigma_{\nu}k^{\nu}$ is negative.
Therefore the smooth solution (\ref{smooth}) for 
Eqs.~(\ref{BGconserv},\ref{BGconserv1}) does not exist.
In this case the only possibility to satisfy the conservation laws
(\ref{BGconserv},\ref{BGconserv1}) 
is to introduce a discontinuity between the fluid 
$T_f$, $\mu_i^f$, $u_f^{\mu}$ and gas
$T_g$, $\mu_i^g$, $u_g^{\mu}$ parameters.
Let us consider this FO shock on t.l. hypersurface in more detail.
To obtain the analytical solution of the problem we restrict
ourself to the case of zero baryonic number $n=0$
and consider massless particles
with the J\"utner distribution function
\begin{equation}\label{jut}
\phi\left(\frac{k_{\mu}u^{\mu}}{T}\right)~
=~\exp\left(-~\frac{k_{\mu}u^{\mu}}{T}\right)~.
\end{equation}
Similar problem, but for the conserved particle number (not the charge!),
was considered in Ref. \cite{3}. 
%%% MYM

Substituting it into both sides of the Eq.~(\ref{BGconserv}),
one finds
that the energy-momentum conservation between the fluid and free particles 
along the t.l. hypersurface
%
%\begin{equation}\label{tmunuconserv}
%d\sigma_{\mu} T^{\mu \nu}_f~=~
%d\sigma_{\mu} T^{\mu \nu}_g~\,\,,
%\end{equation}
%
acquires the form:
\begin{eqnarray}\label{conserv}
& &~ d\sigma_{\mu} T^{\mu \nu}_f \equiv
d\sigma_{\mu} \int \frac{d^3k}{k^{\ro}}~k^{\mu}k^{\nu}~
\phi\left(\frac{k_{\nu}u^{\nu}_f}   
{T_f}\right)~
= \nonumber \\ 
&=&~
d\sigma_{\mu} T^{\mu \nu}_g~\equiv 
d\sigma_{\mu}\int \frac{d^3k}{k^{\ro}}~k^{\mu}k^{\nu}~
\phi\left(\frac{k_{\nu}u^{\nu}_g}
{T_g}\right)~
\theta (d\sigma_{\lambda}k^{\lambda})~.
\end{eqnarray}
The energy-momentum tensor $T^{\mu \nu}_f$ in 
 Eq.~(\ref{conserv}) has the form of Eq.~(\ref{tmunu})
with the functions $\varepsilon(T_f) =\varepsilon_f$ and 
$p(T_f)=p_f$ 
given by 
\begin{equation}\label{peps}
\varepsilon(T)~=~3 p(T) ~=~
\frac{1}{2\pi^2}\int_0^{\infty}k^2dk~k~
~\exp \left(- \frac{k}{T}\right) ~=~
\frac{3}{\pi ^2}~T^4~.
\end{equation}

In the rest frame of 
the FO front, i.e. in 
the Lorentz frame where $d\sigma_{\mu}$ becomes equal to $(0,dt)$, the
$T^{01}_g$ and $T^{11}_g$ components of the gas energy-momentum tensor
in Eq.~(\ref{conserv}) can be rewritten then in the form
\begin{equation}\label{tmunugas}
T^{01}_g ~=~ (\varepsilon ^*_g + p^*_g)u^0_gu^1_g~,~~~~
 T^{11}_g~=~(\varepsilon^*_g + p^*_g) u^1_gu^1_g~ + p^*_g ~.
\end{equation}
Note that  such a representation can be given only for the $T^{\mu\nu}_g$  
components which enter the energy-momentum conservation (\ref{conserv}). 
This form 
coincides with that of Eq.~(\ref{tmunu}),
but with effective values of $\varepsilon ^*_g$
and $p^*_g$ which are found to be equal to
\begin{equation}\label{epsef}
\varepsilon_g^* ~=~\varepsilon(T_g)~\frac{(1+v_g)^2}{4v_g}~,~~~~
p^*_g ~= ~p(T_g)~\frac{(1+v_g)^2(2-v_g)}{4}~,
\end{equation}
where $v_g$ is the velocity parameter of the gas in
the FO shock rest frame.
The functions $\varepsilon$ and $p$ in the right hand side
of Eqs.~(\ref{epsef}) are given by Eq.~(\ref{peps}).

Due to the same formal structure of $T^{01}_g , T^{11}_g$
(\ref{tmunugas}) and $T^{01}_f , T^{11}_f$ (\ref{tmunu})
one obtains the solution of Eq.~(\ref{conserv})
\begin{equation}\label{frvel}
v_{f}^{2} = \frac{(p_f - p_{g}^*)(\varepsilon_g^* + p_{f})}
{(\varepsilon_f - \varepsilon_{g}^*)(\varepsilon_{f} + p_g^*)} \; , \; \;
~ ~ v_g^{2} = \frac{(p_f-p^*_g)(\varepsilon_{f} + p^*_g)}
{(\varepsilon_f - \varepsilon^*_{g})(\varepsilon^*_g + p_{f})}~, 
\end{equation}
which is similar to Eq.~(\ref{slvel}). Note, however,
that in contrast to 
Eq.~(\ref{slvel}) 
the values
of $\varepsilon_g^*$ and 
$p_{g}^*$ in Eq.~(\ref{frvel})
are not just the thermodynamical quantities, but 
depend also on $v_g$. 

It should be emphasized that Eq.~(\ref{tmunugas}) 
can also be introduced
for the distribution functions of massive and charged particles \cite{6}.
In the latter case the energy-momentum conservation leads to 
the familiar expressions (\ref{frvel}) for the effective energy density and 
pressure, 
and the charge conservation
leads to the TA equation (\ref{TA}) or (\ref{nTA}) for the effective charge
density.

We fix the FO hypersurface $\Sigma_f$ by the condition $T_g=const$.
The analytical solution 
of Eq.~(\ref{frvel}) can be presented then in the form
\begin{equation}\label{vgtf}
v_g~=~\frac{9v_f^2 - 8v_f + 3}{3v_f^2 + 1}~,~~~~
R~\equiv~ \left(\frac{T_f}{T_g}\right)^4 ~=~
 \frac{4(3v_f^2 - 2v_f +1)^2 (v_f +1)}{(3v_f -1)(3v_f^2 +1)^2}~.
\end{equation}
It is instructive to compare these results with shock-wave
solution (\ref{slvel}) for the same ideal gas EoS (\ref{peps}).
We fix the temperature $T_g$ of the final state in s.l. shock wave 
(\ref{slshock}) and consider $v_g$ and $T_f$ dependence on $v_f$: 
\begin{equation}\label{vgtfshock}
v_g~=~\frac{1}{3v_f}~,~~~~
R~\equiv~ \left(\frac{T_f}{T_g}\right)^4 ~=~
\frac{3 (1-v_f^2)}{9v_f^2 - 1}~.
\end{equation}

Figs. 2 and 3 show the dependences of $v_g$ and $T_f$ on $v_f$ 
for the fixed value of $T_g$ in the final state, both 
for the FO shock 
(\ref{vgtf}) and for the normal shock wave (\ref{vgtfshock}). The 
kinematical restrictions on the gas velocity 
give the same value of the minimal fluid velocity in both  shock 
transitions,
$(v_f)_{min}= p_f/\varepsilon_f$. It equals to $1/3$
for the considered ideal gas EoS (\ref{peps}) for the fluid.
Figs.~2 and 3 indicate that at low values of the fluid velocity,
$v_f < 1/\sqrt{3}$,
the behavior of $v_f$ and $T_f/T_g$ for the FO shock
(\ref{vgtf}) and for the normal shock wave (\ref{vgtfshock})
is quite similar. The value of $v_f = 1/\sqrt{3}$ corresponds
to the speed of sound,
$c_s=1/\sqrt{3}$,
 in the system with ideal gas EoS (\ref{peps}).
According to the requirements given by
Eq.~(\ref{sound}) the shock 
transitions at $v_f< c_s$ are mechanically unstable.   
Mechanically stable solutions at $v_f> c_s$ for normal
shock waves and FO shocks have qualitatively different
behavior. For the stable normal shock wave one has
$v_g<v_f$ and $T_f<T_g$ (see Figs.~2, 3), i.e. only compression
normal shock wave 
transitions '{\it f}~'$\rightarrow$'{\it g}~' would be stable.
For the FO shock transitions we find a
completely different behavior, $v_g>v_f$ and $T_f>T_g$,
illustrated in Figs.~2, 3.
In contrast to the normal shock waves, only 
the rarefaction FO shock transitions 
'{\it f}~'$\rightarrow$'{\it g}~' are stable. 
This result is in agreement with an intuitive physical picture
of the FO as a rarefaction process.

The decompression degree in the FO shock can be estimated by the
energy density ratio on both side of the discontinuity.
In Appendix B of  Ref. \cite{6}
the energy density of the gas was found
according to the Landau-Lifshitz definition of the collective velocity 
\cite{groot}.
In order to find the desired quantity it is necessary to 
diagonalize the energy-momentum tensor of the gas
by the Lorentz transformation.
Zero-zero component  of the obtained tensor is  
the energy density.
It is found to be  smaller than  the effective energy density
$\varepsilon_g^*$ and is given by 
\begin{equation}\label{realeps}
T_{g}^{00} (v_g, T_g) \biggl|_{L.L.}  =  \varepsilon\left(T_g\right)
\lp 1 - \frac{\D}{2}\rp^2
~\frac{ \hspace*{ 0.1cm}
\lp  \,\,\sqrt{4+ 2 \D+ \D^2} + 1 + \D \rp
}
{ 3
} \,\,,
\end{equation}
where $\D \equiv 1 - v_g$ and $v_g$ is defined by Eq. (\ref{vgtf}).
The magnitude of decompression in the FO shock is presented in Fig.~4. 
It reaches the  maximal value of about 2 at the edge of  
of the stability region, $v_f=1/\sqrt{3}$.
Therefore, the quantitative change 
of the hydrodynamical values in the FO shock 
is not negligible and it might be manifested in the various applications. 

The entropy flux of the fluid is given by 
\begin{equation}\label{entropyf}
s_f^{\mu}d\sigma_{\mu}~=~s_fu_f^{\mu}d\sigma_{\mu}~,
\end{equation}
where $s_f=(\varepsilon_f+p_f)/T_f$ is the fluid
entropy density. 
Similar expression is valid for the final 'gas' state
in the case of normal shock waves.
The entropy production in the normal shock waves is given by the
formula
\begin{equation}\label{entropy1}
\frac{s_g^{\mu} d\sigma_{\mu} }{s_f^{\mu} d\sigma_{\mu} }~ =~
\frac{1}{3^{3/4} v_f} 
\left[ \frac{9 v_f^2 - 1}{1 - v_f^2} \right]^{1/4}\,\,,
\end{equation}
shown in Fig.~5.
The entropy flux of the gas with the cut-off distribution
function is given by
\cite{6}:
\begin{equation}\label{entropyg}
s_g^{\mu}d\sigma_{\mu}~=~
\int \frac{d^3k}{k^{\ro}}~k^{\mu}d\sigma_{\mu}~
\phi\left(~\frac{k_{\nu}u^{\nu}_g}
{T_g}\right)~
\left[~1~-~\ln ~\phi \left(~\frac{k_{\nu}u^{\nu}_g}{T_g}\right)\right]
~\theta (d\sigma_{\lambda}k^{\lambda})~.
\end{equation}
The entropy production in the FO shocks is calculated then as
\begin{equation}\label{entropy2}
\frac{s_g^{\mu} d\sigma_{\mu} }{s_f^{\mu} d\sigma_{\mu} }~ =~ 
\frac{1}{2 v_f} 
\left[ \frac{(3 v_f^2 + 1)^2(1 - 3 v_f)}{(1 + v_f)} \right]^{1/4}\,\,.
\end{equation}
The maximal entropy production in the FO shock
(\ref{entropy2}) corresponds to $v_f=c_s=1/\sqrt{3}$
and can be considered as an analog of the Chapman--Jouguet
point.
One has to note that the Chapman--Jouguet point 
of the FO shock 
is also the boundary of the mechanical stability. It is similar 
to the case of the usual shocks \cite{Go87,Bug88, Bug89, Bug91, rshocks}. 

The initial $(\varepsilon_f,p_f)$ values in the FO shock belong
to the straight line $p=\varepsilon/3$ in the $\varepsilon$--$p$ plane.
The respective final effective values
$(\varepsilon ^*_g,p^*_g)$ are defined by the energy-momentum
conservation (\ref{tmunugas}).
For the fixed FO temperature ($T_g=const$)
the
possible values of $(\varepsilon_f,p_f)$ and the corresponding
values of $(\varepsilon ^*_g,p^*_g)$ are shown in Fig.~6.
The point O in Fig. 6 corresponds to 
$v_f\rightarrow 1$. It follows that $v_g\rightarrow 1$
as well and the non-cut distribution 
function of the gas becomes identical to the fluid one. 
If the fluid velocity decreases, 
the gas velocity decreases also.
The difference between the fluid and the gas
becomes then more pronounced.
This leads to a small growth of 
the gas effective energy (on the arc OB ) and
and an essential growth of the fluid energy density (on the interval OA).
The point A is the Chapman-Jouguet point, which corresponds to the maximal 
value of the effective energy density of the gas.
The unstable final states also belong to the same arc BO, but
they correspond to the opposite direction of changes. 
The point A is the mechanically stable initial state of the fluid with
maximal energy density. 
The mechanically unstable initial states  
are located to the right hand side of the point A.

Curve AOB in Fig.~6  is not a usual TA.
The principal difference is that the gas temperature
is fixed by the FO condition and there are
no states with intermediate temperatures
between the given initial and final ones.
However, the mechanical stability condition (\ref{mechstab})
corresponds exactly to the case of usual shocks and TA.
The critical curve (\ref{pcr}) for points A and B in Fig.~6
is a tangent to the line OA at point A.
The stable FO shock transitions satisfy the mechanical stability
criterion. For the unstable FO shocks the critical curve
has the intersection
point with the line $p=\varepsilon/3$ within the interval OA.
Suppose there is such an intersection point D
for the FO shock transition from the initial state C
to the final state E of the arc BO in Fig.~6.
Then Fig.~5 demonstrates  that entropy
production  on the DE interval is smaller
than that on the interval AB. Moreover, the rarefaction  shock transition
C$\rightarrow$D  is unstable because it connects the two states of
the thermodynamically normal matter. Therefore in the C$\rightarrow$D
transition the
entropy decreases. Combining both results together, one finds
that the  tangent point A of the critical curve corresponds to the
maximum of the entropy production, i.e., it is a Chapman-Jouguet point.

\vspace{0.5cm}
\noindent
{\bf 5. Summary}

\vspace{0.3cm}
\noindent
In the present paper 
the particle freeze-out
in relativistic hydrodynamics
 was considered.
For time-like parts of the FO hypersurface it
is described by the cut-off distribution function
and particle emission looks as a discontinuity
in the hydrodynamical motion (FO shocks). The connection
of these FO shocks to the normal shock waves 
in the relativistic hydrodynamics
has been analyzed. 

The FO shock transition of the fluid to the gas
of free particles with the cut-off distribution function
is mechanically stable if the fluid velocity $v_f$ 
in the rest frame of the FO shock front is larger
then the speed of sound $c_s$ inside the fluid.
This condition coincides with the mechanical stability criteria
for the normal shock waves. In this case the particle emission
does not influence the fluid evolution
and can be selfconsitently included in the
relativistic hydrodynamical models.
We have used only the consequence  (21) of the mechanical stability condition. 
Its complete  formulation for the t.l. F.O. requires further 
studies.

 Analytical solution for the FO shocks are obtained
in the case of the ideal gas EoS (\ref{peps}). 
In contrast to the normal shock waves, 
FO shocks are found to be the rarefaction shocks:
$T_g<T_f$ and also $p_g^*<p_f$, $\varepsilon_g^*<\varepsilon_f$.
For the normal shock waves such behavior corresponds
to the matter with anomalous thermodynamical properties.  
The entropy increase (thermodynamical stability)
takes place for the rarefaction FO shocks, but the entropy
production (\ref{entropy2}) appears to be quite small.

The FO procedure across a time-like
hypersurface considered in this paper
could have rather wide field of applications.
It is necessary for a correct transformation
of hydrodynamical quantities into the spectra of produced particles. 
One example is
the model of A+A collisions which combines hydrodynamics for the early
deconfined stage of the reaction with a microscopic 
non-equilibrium model for the later hadronic
stage (see Ref.~\cite{Ba:99}).

\vspace{0.5cm}
\noindent
{\bf Acknowledgments}

\vspace{0.3cm}
\noindent
Authors thank D.H. Rischke and Granddon Yen for useful discussions.
M.I.G. acknowledges the financial support of DFG, Germany.
K.A.B. is grateful to the Alexander von Humboldt Foundation for  
the financial support.

\newpage

%\section*{References}

\newpage
\input mvplots.tex

\end{document}

%% file: mvplots.tex
\newpage

\begin{figure}
\mbox{\hspace*{1.0cm}\psfig{figure=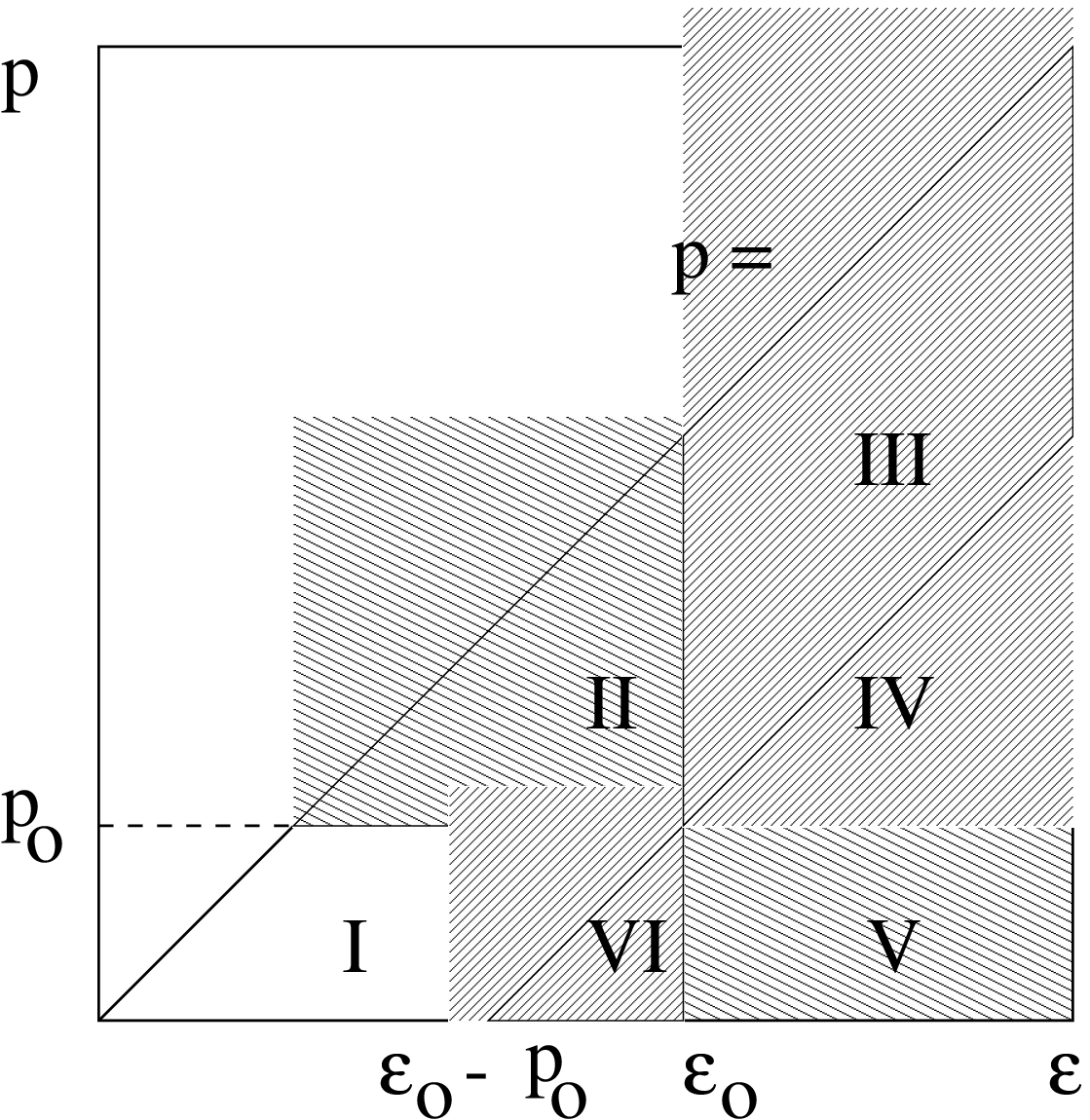,height=14cm,width=14cm}}\\

\vspace{0.5cm}

{\bf Fig. 1.} Possible final states in the $\varepsilon -p$ plane for
shock transitions from the initial state $(\epsilon_{\ro} ,p_{\ro}$).
I and IV are the physical regions for s.l. shocks, III and VI
for t.l. shocks. II and V are unphysical regions for both
types of shocks. Note, that only states with $p\le \varepsilon$ are
possible for any physical equation of state in relativistic theory.

\end{figure}

\clearpage
\newpage

\begin{figure}
\mbox{\psfig{figure=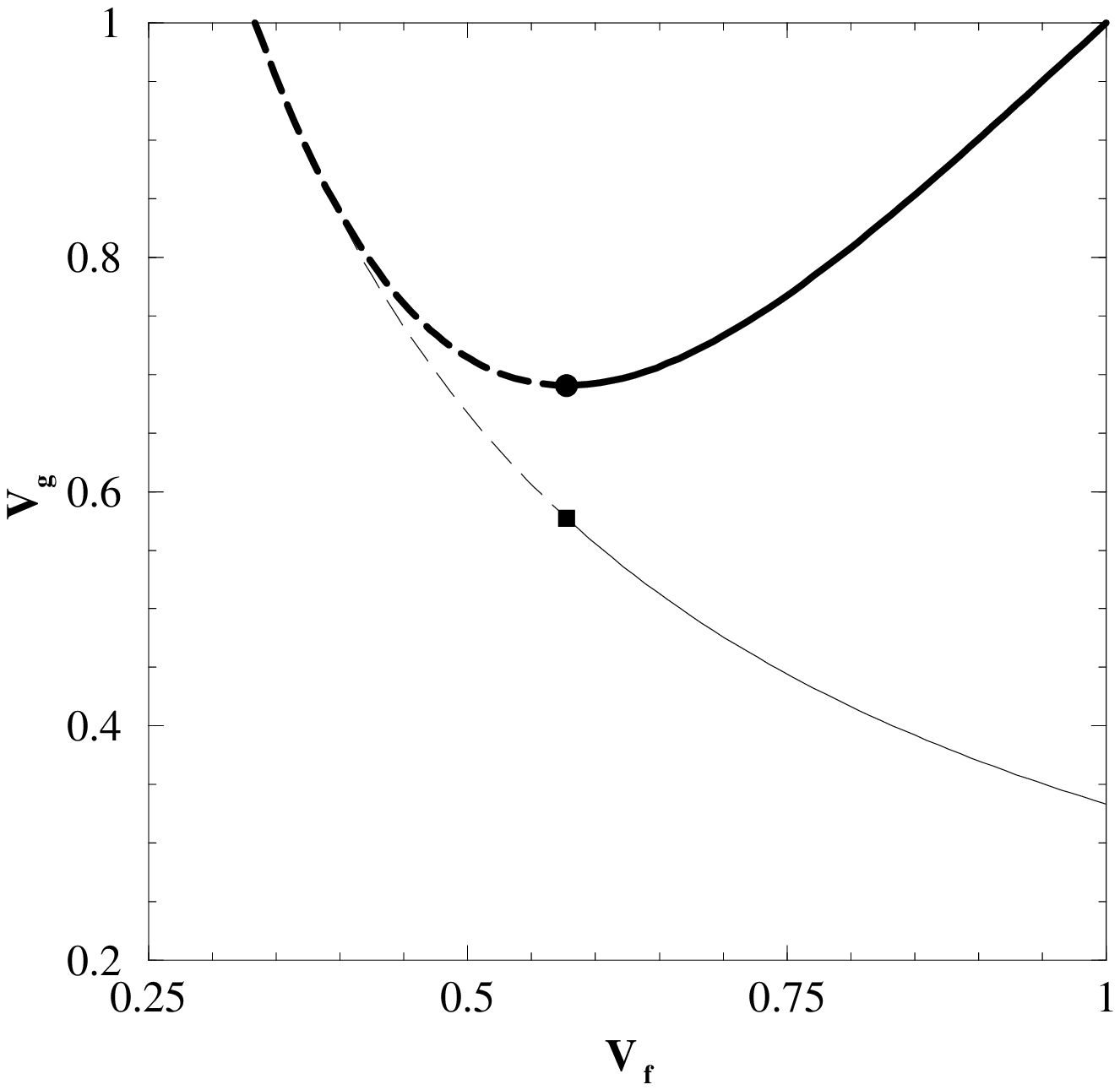,height=15cm,width=17cm}}\\
{\bf Fig. 2.} Gas velocity $v_g$  
as a function of
the fluid velocity $v_f$ in the rest frame of the  
shock front.
The dashed lines represent mechanically unstable transitions,
whereas the solid lines show the mechanically stable FO shocks
and normal shock waves.
Thin lines correspond to a normal shock wave (Eq. (\ref{vgtfshock})) and the      thick
ones correspond to the \fos \mbox{(Eq. (\ref{vgtf}))}.
The circle has the coordinates $(1/\sqrt{3};  
3 - 4/\sqrt{3})$
and the square has the coordinates 
$(1/\sqrt{3}; 1/\sqrt{3})$.

\end{figure}

\newpage

\begin{figure}
\mbox{\psfig{figure=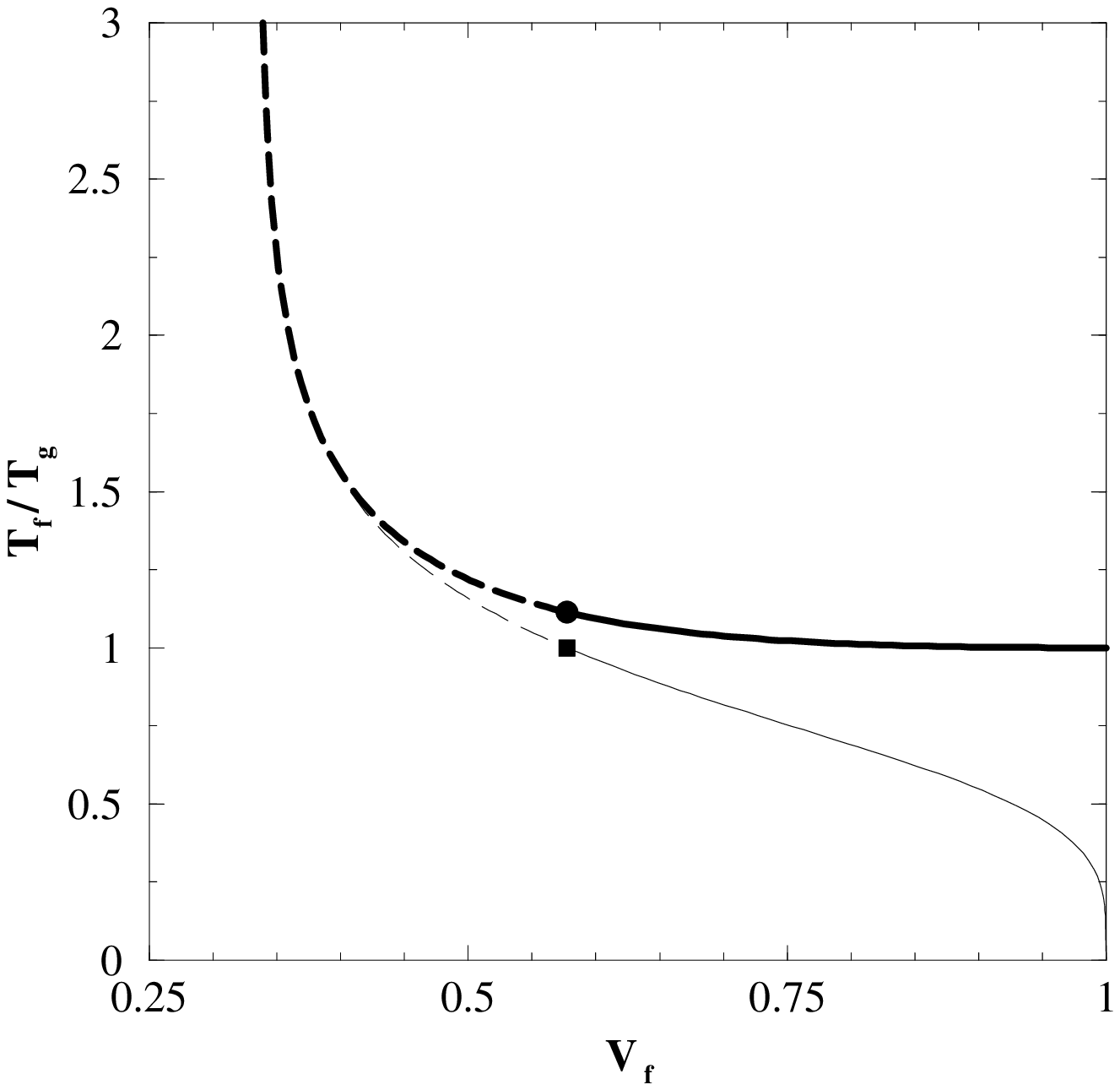,height=15cm,width=17cm}}\\
{\bf Fig. 3.} Ratio of the temperatures on the both
sides of the shock front, $T_f/T_g$, as a function of
the fluid velocity $v_f$ in the rest frame of the
shock front.
The legend corresponds to the Fig. 2.
The circle has the coordinates 
$\left[1/\sqrt{3}; (8/(3 \sqrt{3}))^{1/4} \right]$
and the square has the coordinates $(1/\sqrt{3}; 1)$.

\end{figure}

\newpage
\begin{figure}
\mbox{\psfig{figure=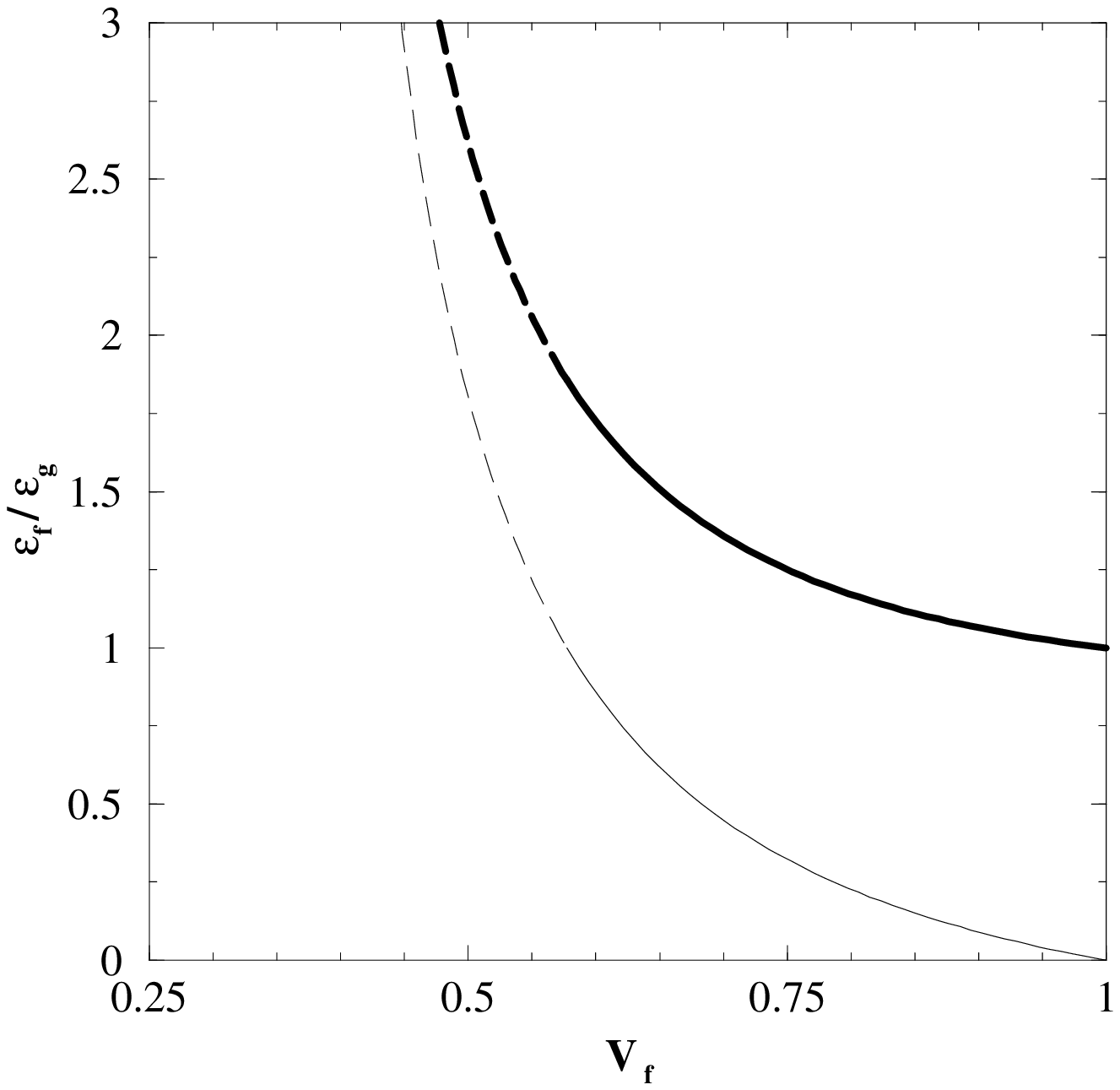,height=15cm,width=17cm}}\\

{\bf Fig. 4.} Ratio of the energy densities on the both
sides of the shock front, as a function of
the fluid velocity $v_f$ in the rest frame of the
\fos\, front.
The dashed lines represent mechanically unstable transitions,
whereas the solid lines show the mechanically stable FO shocks
and normal shock waves.
Thin lines correspond to a normal shock wave  and the
thick ones correspond to the \fos. 
The energy density $\varepsilon_g$
is given by Eq. (\ref{peps})
for the normal shock wave
and for  the gas of free particles 
after the FO shock
it is $T_{g}^{00} (v_g, T_g) \biggl|_{L.L.}$ 
defined by  \mbox{ Eq. (\ref{realeps})}.  

\end{figure}

\newpage
\begin{figure}
\mbox{
\hspace*{3.0cm}\psfig{figure=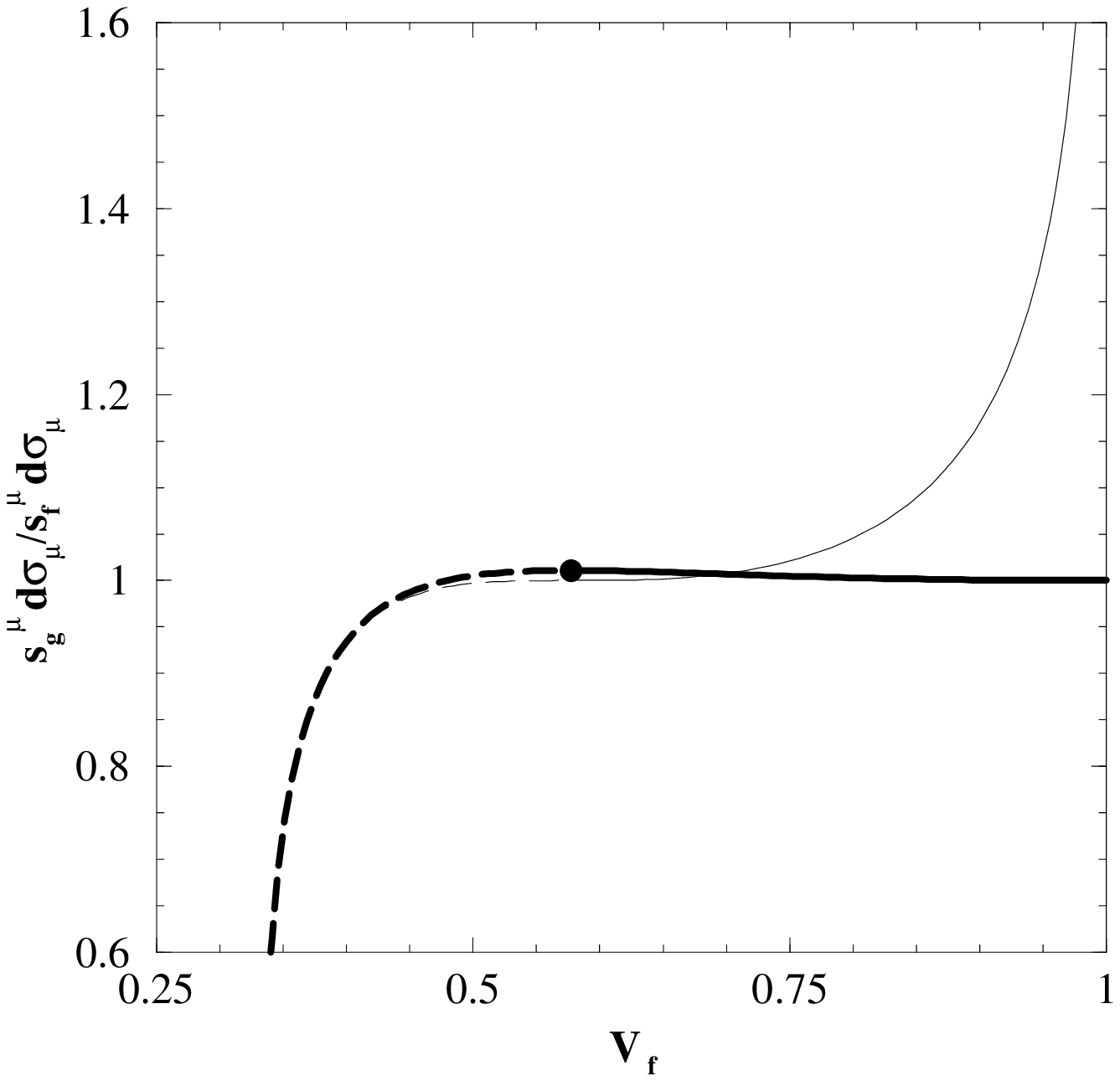,height=9cm,width=9cm} 
}

\vspace*{2.0cm}

\noindent
\mbox{
\hspace*{3.0cm}\psfig{figure=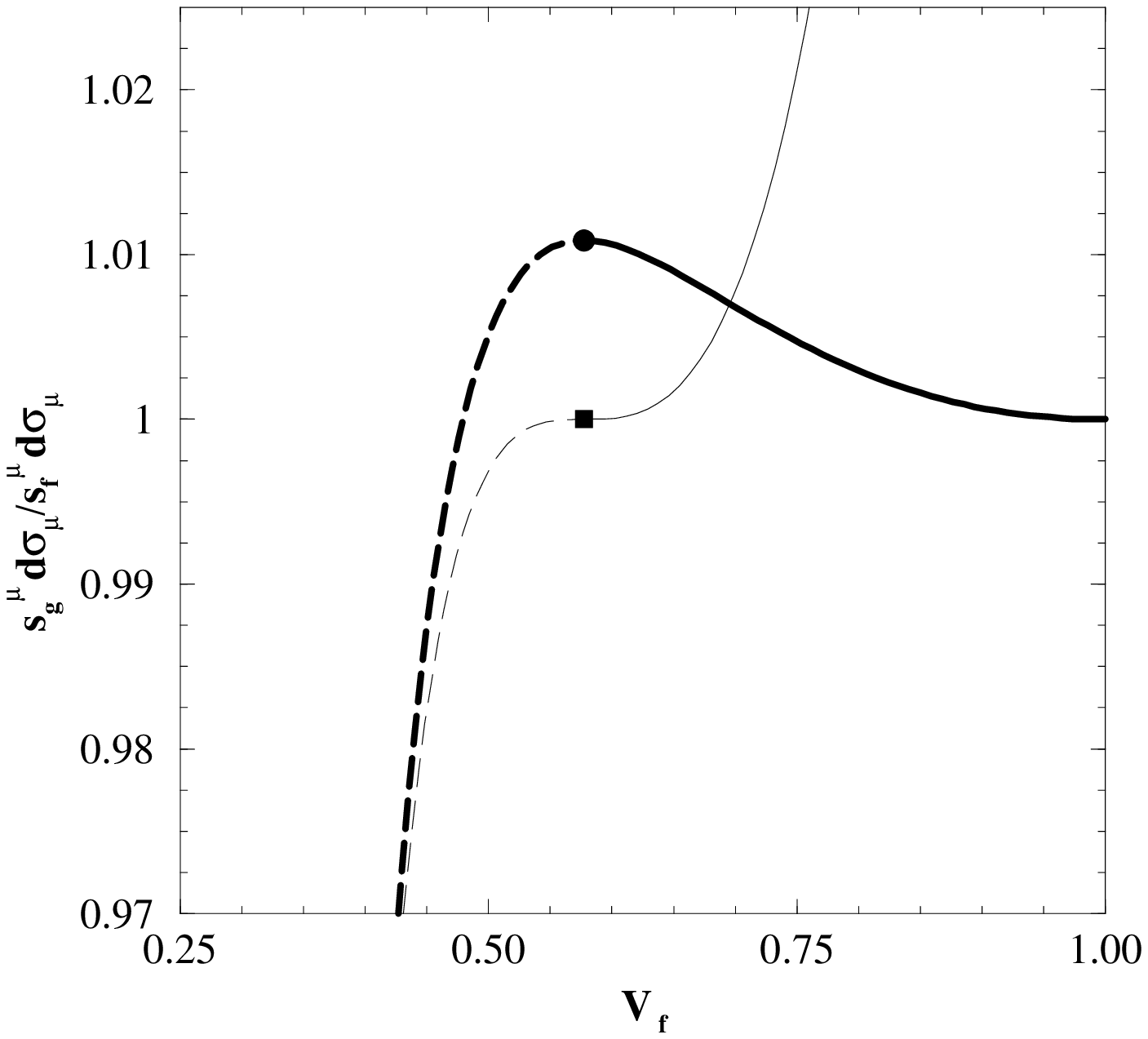,height=9cm,width=9cm}
}

{\bf Fig. 5.} Ratio of the entropies on the both
sides of the shock front, as a function of
the fluid velocity $v_f$ in the rest frame of the
shock front.
The legend corresponds to the Fig. 2.
The upper and lower panels differ by the vertical scale.
The circle corresponds to the maximal entropy 
(analog of the Chapman-Jouguet point)
in the
\fos and it has the coordinates
$\left[1/\sqrt{3};   
\left(\frac {9 (3 - \sqrt{3}) }{4 (\sqrt{3} + 1)}\right)^{1/4} \right]$
and the square has the coordinates $(1/\sqrt{3}; 1)$.

\end{figure}

\newpage
\begin{figure}
\hspace*{-1.5cm}\mbox{\psfig{figure=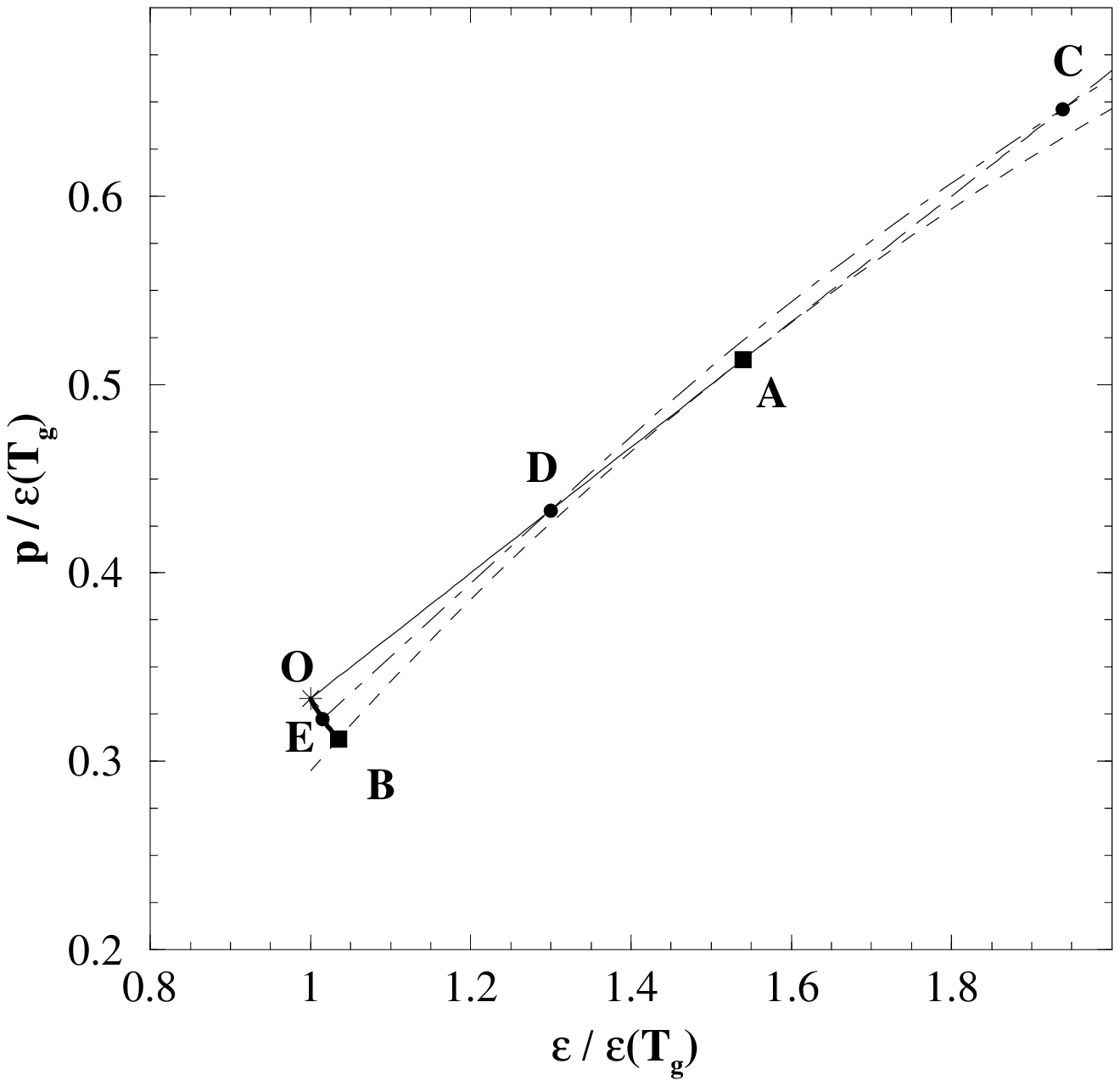,height=18cm,width=21cm}}\\

{\bf Fig. 6.} Possible initial and  final states  in the
$\varepsilon-p$ plane for
the freeze-out shock transitions.
Initial fluid states are on the straight line $p=\varepsilon/3$.
Each initial state on  OA interval corresponds to
the one point on  curve OB of the final gas states in the 
FO shock (see text also).
The initial fluid states of the unstable
FO shock transitions are shown 
by the long dashed line.
The critical curve $p_{cr}(\varepsilon)$
for A$\rightarrow$B FO shock (short dashed line) is a tangent
to the line $p=\varepsilon/3$ at the
point A which is an analog of the
Chapman-Jouguet point. The critical curve $p_{cr}(\varepsilon)$
for C$\rightarrow$E FO shock (dashed-dotted line) crosses
the line $p=\varepsilon/3$ at the point D.

\end{figure}

%%%%%%%%%%%%%%%%%%%%%%%%%%%%%%%%%%%%%%%
%%%%%%%%%%%%%%%%%%%%%%%%%%%%%%%%%%%%%%%